\documentclass[12pt]{article}
\pdfoutput=1
\usepackage{amssymb}
\usepackage{amsmath}
\usepackage{verbatim} 
\usepackage{graphicx,epsfig}
\usepackage{epsfig}

\newcommand{\be}{\begin{equation}}
\newcommand{\ee}{\end{equation}}
\newcommand{\bea}{\begin{eqnarray}}
\newcommand{\eea}{\end{eqnarray}}
\newcommand{\beas}{\begin{eqnarray*}}
\newcommand{\eeas}{\end{eqnarray*}}
\newcommand{\nn}{\nonumber}

\begin{document}
\numberwithin{equation}{section}

\baselineskip 14 pt
\parskip 0 pt 

\begin{titlepage}

\begin{center}

\vspace{5mm}

{\Large \bf  Boundary Terms and Junction Conditions for the DGP $\pi$-Lagrangian and Galileon}

\vspace{5mm}

Ethan Dyer\footnote{esd2107@columbia.edu} and Kurt Hinterbichler\footnote{kurth@phys.columbia.edu}

\vspace{2mm}

{\small \sl Institute for Strings, Cosmology and Astroparticle Physics} \\
{\small \sl and Department of Physics} \\
{\small \sl Columbia University, New York, NY 10027 USA}

\end{center}

\vskip 1.0 cm

\noindent
In the decoupling limit of DGP, $\pi$ describes the brane-bending degree of freedom.  It obeys second order equations of motion, yet it is governed by a higher derivative Lagrangian.  We show that, analogously to the Einstein-Hilbert action for GR, the $\pi$-Lagrangian requires Gibbons-Hawking-York type boundary terms to render the variational principle well-posed.  These terms are important if there are other boundaries present besides the DGP brane, such as in higher dimensional cascading DGP models.  We derive the necessary boundary terms in two ways.  First, we derive them directly from the brane-localized $\pi$-Lagrangian by demanding well-posedness of the action.  Second, we calculate them directly from the bulk, taking into account the Gibbons-Hawking-York terms in the bulk Einstein-Hilbert action.  As an application, we use the new boundary terms to derive Israel junction conditions for $\pi$ across a sheet-like source.  In addition, we calculate boundary terms and junction conditions for the galileons which generalize the DGP $\pi$-lagrangian, showing that the boundary term for the $n$-th order galileon is the $(n-1)$-th order galileon.

\end{titlepage}
\setcounter{footnote}{0}

\section{Introduction and results\label{Intro}}
\ \ \ \ \
The Dvali-Gabadadze-Porrati (DGP) model is a higher-dimensional brane-world model first introduced in \cite{Dvali:2000hr}.  The defining feature of DGP is an Einstein-Hilbert term localized on the brane, in addition to the one in the bulk.  DGP has its problems (see \cite{Gregory:2008bf} for a sample), but we still find it interesting and worth studying due to the fact that quantum corrections will generically induce a DGP term in any brane-world setup, and thus the problems can be used to place constraints on possible UV completions \cite{Adams:2006sv,Hinterbichler:2009kq}, or may be remedied by going to higher codimension or cascading DGP setups \cite{deRham:2007xp}.  We consider the case where there is an $(n-1)$-dimensional brane in an $n$-dimensional bulk\footnote{Here $X^A$, with $A,B,\cdots=0,1,2,3,\ldots,n-1$ are the $n$-dimensional bulk coordinates, $G_{AB}(X)$ is the $n$-dimensional metric, and $M_n$ is the $n$-dimensional Planck mass.  $x^\mu$, with $\mu,\nu,\ldots=0,1,2,3,\ldots,n-2$ are the $(n-1)$-dimensional brane coordinates, $g_{\mu\nu}(x)$ is the $(n-1)$-dimensional brane metric given by inducing the $n$-dimensional metric $G_{AB}$ onto the brane, and $M_{n-1}$ is the $(n-1)$-dimensional Planck mass.  $S_M$ is the matter action, which we imagine to be localized to the brane.  We call the $n-1$ coordinate $y$, and choose coordinates such that the DGP brane lies at $y=0$.  We denote the bulk volume ${\cal V}$, and the DGP brane ${\cal B}_y$.}, 
 \be S={M_n^{n-2}\over 2}\int_{\mathcal{V}} d^nX\sqrt{-G}R(G)+{M_{n-1}^{n-3}\over 2}\int_{\mathcal{B}_{y}} d^{n-1}x\sqrt{-g}R(g)+S_M.\ee
  
One is often interested in integrating out the bulk modes to find an effective $(n-1)$-dimensional description of the system \cite{Luty:2003vm}.  It is found that in a certain limit, called the decoupling limit, 
the effective boundary theory reduces to linear gravity coupled to the matter stress tensor, along with the following non-linear action for a scalar degree of freedom\footnote{All the derivatives here are $(n-1)$-dimensional.  The metric signature convention is mostly plus.  The factor of 3 normalizing the scalar kinetic term is conventional.}, 
\be \label{piLagrangian} S_\pi=\int_{\mathcal{B}_{y}} d^{n-1}x\ -3(\partial\pi)^2-{1\over 2}\lambda(\partial\pi)^2\square\pi+\pi T,\ee
where $T$ is the trace of the matter stress tensor scaled by appropriate constants, and $\lambda$ is a dimensionful coupling constant, which reflects the strong coupling scale of the theory (the precise value of $\lambda$ does not concern us, but it can be found in terms of $M_n$ and $M_{n-1}$).  This is known as the $\pi$ Lagrangian\footnote{In addition, this Lagrangian (\ref{piLagrangian}) is invariant, up to a total derivative, under the internal galilean symmetry  $ \pi\rightarrow \pi+c+b_\mu x^\mu,$ where $c$ and $b_\mu$ are constants.  It is an example of a larger class of Lagrangians with this symmetry, dubbed galileons \cite{Nicolis:2008in}.}, first obtained in \cite{Luty:2003vm}, and further studied in \cite{Nicolis:2004qq}.  The field $\pi$ is a certain configuration of bulk fields, namely, 
\be \label{piconfig} \tilde N=\partial_y\Pi,\ \ \ \tilde N_\mu=\partial_\mu \Pi,\ \ \ h_{\mu\nu}=0,   \ee
where $\Pi(x,y)=e^{-y\Delta}\pi(x)$, $\Delta=\sqrt{-^{(n-1)}\square}$.  Here $\tilde N(x,y)$, $\tilde N_\mu(x,y)$ and $h_{\mu\nu}(x,y)$ are the deviations of the $n$-dimensional lapse, shift (both with respect to the $y$ direction) and brane metric about a flat background.  The cubic part of the $\pi$ Lagrangian comes from expanding the bulk Einstein-Hilbert term to third order in the lapse and shift and evaluating on the configuration (\ref{piconfig}). 

\begin{figure}[h!]
\begin{center}
\includegraphics[height=4in]{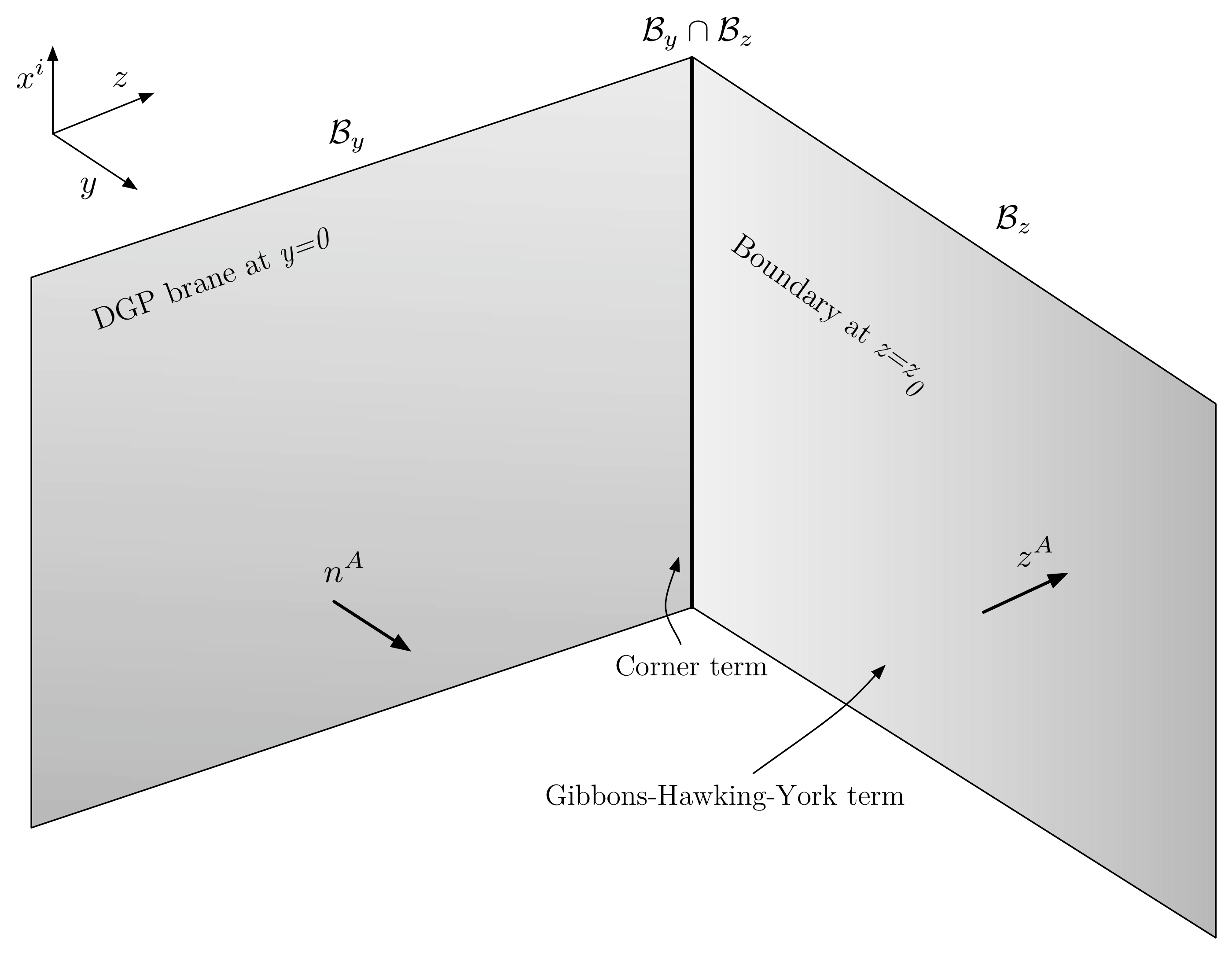}
\caption{Setup}
\label{boundaries}
\end{center}
\end{figure}

We are interested in the situation in which the DGP brane itself has some kind of $n-2$ dimensional boundary to it, for whatever reason.  For example, suppose the bulk has, in addition to the DGP brane, an $(n-1)$-dimensional boundary located at $z=z_0$, where $z$ is one of the spatial coordinates transverse to $y$.   Calling the DGP brane at $y=0$ by ${\cal B}_y$ and the boundary at $z=z_0$ by ${\cal B}_z$, the intersection ${\cal B}_y\cap {\cal B}_z$ is a $(n-2)$-dimensional boundary for the DGP brane.  $x^i$ will be the $(n-2)$ coordinates along the intersection.  See figure \ref{boundaries} for the setup.    

There are any number of reasons such a boundary may be present physically.   Examples include intersecting brane models in string theory, and cascading DGP \cite{deRham:2007xp}.   However, even if a boundary such as ${\cal B}_z$ is not present physically, we are still interested in imagining it off at infinity, representing spatial boundary conditions for the fields.  These considerations are necessary both for the purposes of rendering the action well-posed, and for making sense of global concepts such as energy and entropy (see \cite{Dyer:2008hb} for discussions of these and other points).  

The field equations stemming from (\ref{piLagrangian}) are second order, so there is only a single degree of freedom.  Accordingly, we many only set one piece of boundary data.  Varying the action (\ref{piLagrangian}), we must keep track of boundary contributions at $z=z_0$.   We will employ a variational principle where we fix the value of $\pi$ on the boundary, which implies that the variation $\delta\pi$ (and hence its tangential derivatives) equal zero on the boundary.  We may not set the normal derivatives $\partial_z\delta\pi$ equal to zero on the boundary \cite{Dyer:2008hb}.  

As it stands, there is a leftover boundary contribution to the variation of (\ref{piLagrangian}), which contains $\partial_z\delta\pi$ and may not be set to zero,
\be\label{boundaryvariation} \delta S_\pi = -\lambda\int_{\mathcal{B}_{y}\cap\mathcal{B}_{z}}d^{n-2}x\ \frac{1}{2}(\partial_{\mu}\pi)^{2}\partial_{z}\delta\pi.
\ee
This variation must be cancelled if the action is to be truly stationary and the variational principle well-posed.  As we'll see, this can only be achieved if the following boundary term is added to the $\pi$-action,
\be \label{piboundaryterms}
   S_B=\lambda \int_{\mathcal{B}_{y}\cap\mathcal{B}_{z}}d^{n-2}x \ \frac{1}{6}(\partial_{z}\pi)^{3} + \frac{1}{2}(\partial_{i}\pi)^{2}\partial_{z}\pi.
   \ee
The variation of (\ref{piboundaryterms}) exactly cancels (\ref{boundaryvariation}). The boundary term (\ref{piboundaryterms}) is analogous to the Gibbons-Hawking-York boundary term that must be added to the Einstein-Hilbert action to render it well-posed \cite{Gibbons:1976ue,York:1972sj,Dyer:2008hb}.  It arises for a similar reason; the Lagrangian is higher order, leading to more derivatives on boundary variations, and yet the field equations are lower order, leading to fewer degrees of freedom for setting boundary variations to zero.  

The bulk DGP action is simply the Einstein-Hilbert action, and is well-posed if the appropriate Gibbons-Hawking-York terms are added to all boundaries.  As such, we should expect to be able to derive a well-posed $\pi$-Lagrangian in the decoupling limit.  We will derive the appropriate boundary terms (\ref{piboundaryterms}) from the bulk. 

To make the bulk Einstein-Hilbert action well-posed, Gibbons-Hawking-York terms must be present on ${\cal B}_y$ and ${\cal B}_z$, and a corner term must be present on the $(n-2)$-dimensional juncture ${\cal B}_y\cap {\cal B}_z$ \cite{Hayward:1993my}.  Upon adding these terms, we will re-derive the $\pi$-action by expanding the bulk action to cubic order in the lapse and shift, as in \cite{Luty:2003vm}, but keeping careful track of all boundary contributions.  We will find that the boundary terms (\ref{piboundaryterms}) come out automatically, as expected from well-posedness of the bulk action.  

Finally, we will use the new boundary term (\ref{piboundaryterms}) to derive, directly from the action principle, junction conditions for the $\pi$ field across a sheet-like source of the form $T(x^i,z)=T(x^i)\delta(z)$, i.e. the analog of the Israel junction conditions of GR.  We will find that the normal derivative, $\delta_z\pi$, of the $\pi$ field should change across the source as 
\be 2\lambda\partial_i^2\pi\Delta\left[\partial_z\pi\right]=-T(x^i).
\ee

\section{Derivation of the boundary term from well-posedness of the $\pi$-action}
\ \ \ \ \
Consider the cubic part of the $\pi$-Lagrangian (\ref{piLagrangian}) as an $(n-1)$-dimensional theory on ${\cal B}_y$, forgetting for the moment that it comes from DGP,  
\begin{eqnarray}\label{piLagrangian2}
S_{\pi}=-\frac{1}{2}\int_{\mathcal{B}_{y}}d^{n-1}x \ (\partial\pi)^{2}\square\pi.
\end{eqnarray}
(We will be working only with the cubic term so we drop the overall constant, $\lambda$.)  Varying the action, we find 
\begin{eqnarray}\label{pivary}
 \delta{S_{\pi}}&=&\int_{\mathcal{B}_{y}}d^{n-1}x \ \left[(\square\pi)^{2}-(\partial_{\mu}\partial_\nu\pi)^{2}\right]\delta\pi\nonumber\\ \label{fullboundaryvariation}
 &-&\int_{\mathcal{B}_{y}\cap\mathcal{B}_{z}}d^{n-2}x \ [\left(\partial_{z}\pi\right)\square\pi - \partial^{\mu}\pi\partial_{z}\partial_{\mu}\pi]\delta\pi +\frac{1}{2}(\partial_{\mu}\pi)^{2}\partial_{z}\delta\pi,
 \end{eqnarray}
 where we have kept terms on the boundary $\mathcal{B}_{y}\cap\mathcal{B}_{z}$.   
 
The field equations are second order despite the fact that the Lagrangian has second derivatives, which cannot be removed with an integration by parts.  This leads to many nice properties \cite{Nicolis:2004qq}.   The fact that the field equations are second order means that there is 
only a single propagating scalar degree of freedom (as opposed to higher order equations, which may describe more than one degree of freedom).  The number of boundary conditions that may be set is equal to the number of degrees of freedom, so we may only set one boundary condition for $\pi$ \cite{Dyer:2008hb}.  We will choose to fix the value of $\pi$ on the boundary.  This implies $\delta\pi=0$ on the boundary, as well as $\partial_i\delta\pi=0$ on the boundary.  Given this, we may not set $\partial_z\delta\pi=0$ on the boundary, as this would be fixing an additional degree of freedom.  Because of the contribution $-\frac{1}{2}(\partial_{\mu}\pi)^{2}\partial_{z}\delta\pi$ to the boundary variation (\ref{fullboundaryvariation}), the action is not stationary.  The action $S_\pi$ is therefore not well-posed when $\pi$ is fixed on the boundary.  

We wish to find a boundary term, $S_{B}$, which when added to $S_\pi$, yields an action $S'_\pi=S_\pi+S_B$, which is well posed.  The variation of $S_B$ must therefore cancel the unwanted term containing $\partial_z\delta\pi$, 
\be\label{wantedvariation} \delta S_B=\int_{\mathcal{B}_{y}\cap\mathcal{B}_{z}}d^{n-2}x\ \frac{1}{2}(\partial_{\mu}\pi)^{2}\partial_{z}\delta\pi.\ee
Consider the terms that are potentially present in $S_B$.   Each must have three factors of $\pi$, and three derivatives, an even number of which must be $\partial_{i}$'s (because they must contract).  Thus there can be either one $\partial_z$ and two $\partial_i$'s or three $\partial_z$'s.  There cannot be more than one z derivative acting on each $\pi$ (because then the variation would contain $\partial_z^2\delta\pi$ or $\partial_z^3\delta\pi$, which cannot be set to zero). This leaves three possible terms, up to a total tangential derivative, which we write with arbitrary coefficients $a$, $b$ and $c$,
 \begin{eqnarray}
 S_B=\int_{\mathcal{B}_{y}\cap\mathcal{B}_{z}}d^{n-2}x \ a\ (\partial_{z}\pi)^{3} + b\ (\partial_{i}\pi)^{2}\partial_{z}\pi+c\ \pi\partial_{i}^2\pi\partial_{z}\pi.
 \end{eqnarray}
 The variation of this yields (setting $\delta\pi=0$ as per our variational principle),
  \begin{eqnarray}
 \delta{S}_B=\int_{\mathcal{B}_{y}\cap\mathcal{B}_{z}}d^{n-2}x \ \left[3a(\partial_{z}\pi)^{2} + b(\partial_{i}\pi)^{2}+c\pi\partial_{i}^2\pi\right]\partial_{z}\delta\pi.
 \end{eqnarray}
 This gives (\ref{wantedvariation}) if and only if $a=\frac{1}{6}$, $b=\frac{1}{2}$, and $c=0$. Therefore the appropriate boundary term is:
  \begin{eqnarray}\label{piboundary2}
 S_B=\int_{\mathcal{B}_{y}\cap\mathcal{B}_{z}}d^{n-2}x \ \frac{1}{6}(\partial_{z}\pi)^{3} + \frac{1}{2}(\partial_{i}\pi)^{2}\partial_{z}\pi.
 \end{eqnarray}
 Of course there is some freedom in the boundary term; the addition of any function of $\pi$ or $\partial_{i}\pi$ does not alter the variation.  (\ref{piboundary2}) is the unique boundary term involving three $\pi$'s and three derivatives, up to a total tangential derivative and terms involving no normal derivatives.   In the next section, we will derive this boundary term $S_B$ directly from the bulk. 
 
\section{Derivation of the boundary term from the bulk}
 \ \ \ \ \
We first set up some notation.  See the appendices of \cite{Dyer:2008hb} for further information and conventions.  The normal vector to ${\cal B}_y$ is denoted $n^A$, and it points in the positive $y$ direction, into the bulk\footnote{Note that this is opposite to the conventions laid out in \cite{Dyer:2008hb}, so we will have to account for this with an explicit minus sign.}.  The extrinsic curvature of ${\cal B}_y$ is denoted $K_{\mu\nu}$, and the induced metric is $g_{\mu\nu}$.  The normal vector to ${\cal B}_z$ is denoted  $z^A$ , and it points in the positive $z$ direction, out of the bulk.  Indices on ${\cal B}_z$ are $a,b,\ldots$, and the induced metric is denoted $\gamma_{ab}$.  The extrinsic curvature of ${\cal B}_z$ is denoted ${\cal K}_{ab}$.  The metric on ${\cal B}_y\cap {\cal B}_z$ is denoted $\sigma_{ij}$.  The extrinsic curvature on ${\cal B}_y\cap {\cal B}_z$ as embedded in ${\cal B}_y$ is denoted $k_{ij}$.  As we'll see, we cannot require $z^A n_A=0$, so the boundaries do not necessarily intersect orthogonally.  

To derive the well-posed action $S'_\pi$ from the bulk, we must start with a bulk action, which is well-posed.  This means the appropriate Gibbons-Hawking-York terms must be added to both the boundaries ${\cal B}_z$ and ${\cal B} _y$  \cite{Gibbons:1976ue,York:1972sj,Dyer:2008hb}.  The Gibbons-Hawking-York term is simply the trace of the extrinsic curvature of the boundary.  In addition, it has been shown that a corner term must be added when there are non-orthogonal junctions between boundaries \cite{Hayward:1993my}.  The corner term Lagrangian is $\arccos(n^{A}z_{A})$, 
and is localized on $\mathcal{B}_{y}\cap\mathcal{B}_{z}$.  These terms are all proportional to $M_n^{n-2}$, because they are there to render the bulk Einstein-Hilbert Lagrangian well-posed.  The brane Einstein-Hilbert term must also be accompanied by a Gibbons-Hawking-York proportional to $M_{n-1}^{n-3}$.  This term is localized on $\mathcal{B}_{y}\cap\mathcal{B}_{z}$ and involves $k$, the trace of the extrinsic of $\mathcal{B}_{y}\cap\mathcal{B}_{z}$ as embedded in ${\cal B}_y$.  The total action is therefore, 

\begin{eqnarray}
S&=&{M_n^{n-2}\over 2}\int_{\mathcal{V}} d^{n}X\sqrt{-G} \ R(G) + M_n^{n-2}\int_{\mathcal{B}_{z}}d^{n-1}x\sqrt{-\gamma} \ \mathcal{K} -M_n^{n-2}\int_{\mathcal{B}_{y}}d^{n-1}x\sqrt{-g} \ K \nn \\
&-&M_n^{n-2}\int_{\mathcal{B}_{y}\cap\mathcal{B}_{z}}d^{n-2}x\sqrt{-\sigma} \arccos(n^{A}z_{A})\nonumber\\
&+&{M_{n-1}^{n-3}\over 2}\int_{\mathcal{B}_{y}}d^{n-1}x\sqrt{-g} \ R(g) + M_{n-1}^{n-3}\int_{\mathcal{B}_{y}\cap\mathcal{B}_{z}}d^{n-2}x \sqrt{-\sigma}\ k.
\end{eqnarray}
Note that the term on ${\cal B}_y$ comes with a minus sign because, opposite to convention, we have set the normal vector $n^A$ pointing into the bulk.  

Following \cite{Luty:2003vm}, we change to ADM variables in the $y$ direction (see the appendices of \cite{Dyer:2008hb} for conventions on the ADM decomposition).  The $n$-dimensional bulk term and $(n-1)$-dimensional Gibbons-Hawking-York terms (dropping an overall factor of $M_n^{n-2}/2$) are 
\begin{eqnarray}
S&=&\int_{\mathcal{V}} d^nX\sqrt{-g} \ N\left[R + K^{2} - K_{\mu\nu}K^{\mu\nu} + 2\nabla_{A}(n^{B}\nabla_{B}n^{A}-n^{A}\nabla_{B}n^{B})\right]\\
 &+& 2\int_{\mathcal{B}_{z}}d^{n-1}x\sqrt{-\gamma} \ \mathcal{K} -2\int_{\mathcal{B}_{y}}d^{n-1}x\sqrt{-g} \ K -2\int_{\mathcal{B}_{y}\cap\mathcal{B}_{z}}d^{n-2}x\sqrt{-\sigma} \arccos(n^{A}z_{A}). \nonumber
\end{eqnarray}
The total derivative in the first line can be simplified by use of Stokes' theorem, keeping in mind that $n^Az_A\not=0$, 
\begin{eqnarray}
&&\int_{\mathcal{V}} d^{n}X\sqrt{-G} \left[2\nabla_{A}(n^{B}\nabla_{B}n^{A}-n^{A}\nabla_{B}n^{B})\right] \\ \nn &&+ 2\int_{\mathcal{B}_{z}}d^{n-1}x\sqrt{-\gamma} \ \mathcal{K} -2\int_{\mathcal{B}_{y}}d^{n-1}x\sqrt{-g} \ K\nonumber\\
&=&2\int_{\mathcal{B}_{z}}d^{n-1}x\sqrt{-\gamma}\left[\mathcal{K} + z_{A}(n^{B}\nabla_{B}n^{A}-n^{A}\nabla_{B}n^{B})\right]\nonumber\\
&&-2\int_{\mathcal{B}_{y}}d^{n-1}x\sqrt{-g}\left[K + n_{A}(n^{B}\nabla_{B}n^{A}-n^{A}\nabla_{B}n^{B})\right].\nonumber
\end{eqnarray} 
The second integral vanishes as $n^{A}\nabla_{B}n_{A} = 0$, $n^{A}n_{A}=1$, and $\nabla_{B}n^{B}=K$. The first term can be further reduced,
\begin{eqnarray}
&&2\int_{\mathcal{B}_{z}}d^{n-1}x\sqrt{-\gamma}\left[\mathcal{K} + z_{A}(n^{B}\nabla_{B}n^{A}-n^{A}\nabla_{B}n^{B})\right]\nonumber\\
&=&2\int_{\mathcal{B}_{z}}d^{n-1}x\sqrt{-\gamma}\left[\mathcal{K} -n^{A}n^{B}\nabla_{B}z_{A}+ n^{B}\nabla_{B}(n^{A}z_{A})-n^{A}z_{A}\nabla_{B}n^{B}\right]\nonumber\\
&=&2\int_{\mathcal{B}_{z}}d^{n-1}x\sqrt{-\gamma}\left[\gamma^{ab}e^{A}_{a}e^{B}_{b}\nabla_{B}z_{A}-n^{A}n^{B}\nabla_{B}z_{A} + n^{B}\nabla_{B}(n^{A}z_{A})-n^{A}z_{A}\nabla_{B}n^{B}\right]\nonumber\\
&=&2\int_{\mathcal{B}_{z}}d^{n-1}x\sqrt{-\gamma}\left[g^{\mu\nu}e^{A}_{\mu}e^{B}_{\nu}\nabla_{B}z_{A} + n^{B}\nabla_{B}(n^{A}z_{A})-n^{A}z_{A}\nabla_{B}n^{B}\right],
\end{eqnarray}
where $e^{A}_{\mu}$ and $e^{A}_{a}$ are bases tangent to ${\cal B}_y$ and ${\cal B}_z$, respectively.   

Bringing back the corner term, the full action is then, 
\be S=S_{\cal V}+S_{{\cal B}_z}+S_{\rm corner},\ee
where,
\begin{eqnarray}\label{threeterms}
\nn S_{\cal V}&=&\int_{\mathcal{V}} d^nX\sqrt{-g} \ N\left[R + K^{2} - K_{\mu\nu}K^{\mu\nu}\right],\\
S_{{\cal B}_z}&=&2\int_{\mathcal{B}_{z}}d^{n-1}x\sqrt{-\gamma}\left[g^{\mu\nu}e^{A}_{\mu}e^{B}_{\nu}\nabla_{B}z_{A} + n^{B}\nabla_{B}(n^{A}z_{A})-n^{A}z_{A}K\right],\nonumber\\
S_{\rm corner}&=&-2\int_{\mathcal{B}_{y}\cap\mathcal{B}_{z}}d^{n-2}x\sqrt{-\sigma} \arccos(n^{A}z_{A}).
\end{eqnarray}
So far, no approximations or expansions have been made.  To derive the cubic $\pi$ term, we expand the action around a flat background $G_{AB}=\eta_{AB}$, where the branes are flat and perpendicular to each other.  We expand the lapse, shift, and spatial metric accordingly,
\be N=1+\tilde N,\ \ \ N_\mu=\tilde N_\mu,\ \ \ \ g_{\mu\nu}=\eta_{\mu\nu}+h_{\mu\nu}.\ee
The $\pi$ cubic term comes from terms in the expansion which are cubic order in the deviations $\tilde N,\tilde N_\mu$, and zeroth order in $h_{\mu\nu}$.  As such, none of the brane terms proportional to $M_{n-1}^{n-3}$ contribute, since they only depend on $h_{\mu\nu}$.  Thus, we must only expand the three terms (\ref{threeterms}).

Once we have the third order expression, we isolate the scalar mode by plugging in 
\be \label{piconfig2} \tilde N=\partial_y\Pi,\ \ \ \tilde N_\mu=\partial_\mu \Pi,   \ee
where $\Pi(x,y)=e^{-y\Delta}\pi(x)$, $\Delta=\sqrt{-^{(n-1)}\square}$.  The result should completely localize to ${\cal B}_y$.  

Before starting, we collect some useful third order expressions,
\begin{eqnarray} \nn
K_{\mu\nu}&=&-\frac{1}{2}(1-\tilde{N}+\tilde{N}^{2})(\partial_{\mu}\tilde{N}_{\nu}+\partial_{\nu}\tilde{N}_{\mu}),\\ \nn
K&=&-(1-\tilde{N}+\tilde{N}^{2})(\partial_{\mu}\tilde{N}^{\mu}),\\ \nn
z_{A}&=&(1-\frac{1}{2}\tilde{N}_{z}^{2} +\tilde{N}\tilde{N}_{z}^{2})\partial_{A}z,\\ \nn
n^{A}z_{A}&=&-\tilde{N}_{z}+\tilde{N}\tilde{N}_z+\frac{1}{2}\tilde{N}_{z}^{3}-\tilde{N}_{z}\tilde{N}^{2},\\  \label{expressions}
\sqrt{-\gamma}&=&1+\tilde{N}+\frac{1}{2}\tilde{N}_{z}^{2}-\frac{1}{2}\tilde{N}\tilde{N}_{z}^{2}.
\end{eqnarray}

\subsection*{Bulk piece, $S_{\cal V}$}
\ \ \ \ \
Start with the terms over the bulk volume,
\begin{eqnarray}
S_{\mathcal{V}}=\int_{\mathcal{V}} d^nX\sqrt{-g} \ N\left[R + K^{2} - K_{\mu\nu}K^{\mu\nu}\right].
\end{eqnarray}
Because we are only keeping zeroth order in metric deviations, we have $\sqrt{-g}=1$, $R=0$. 
With this the bulk term is,
\begin{eqnarray}
S_{\mathcal{V}}&=&\int_{\mathcal{V}} d^nX \ -\tilde{N}\left[(\partial_{\mu}\tilde{N}^{\mu})^{2} -\frac{1}{2}\partial_{\mu}\tilde{N}_{\nu}(\partial^{\mu}\tilde{N}^{\nu}+\partial^{\nu}\tilde{N}^{\mu})\right]\nonumber\\
&=&\int_{\mathcal{V}} d^nX \ \partial_{\mu}\tilde{N}\left[\tilde{N}^{\mu}\partial_{\nu}\tilde{N}^{\nu} -\frac{1}{2}\tilde{N}_{\nu}(\partial^{\mu}\tilde{N}^{\nu}+\partial^{\nu}\tilde{N}^{\mu})\right] +\frac{1}{2}\tilde{N}\left[\tilde{N}^{\mu}\partial_{\mu}\partial_{\nu}\tilde{N}^{\nu}-\tilde{N}_{\mu}\square\tilde{N}^{\mu}\right]\nonumber\\
&&-\int_{\mathcal{B}_{z}}d^{n-1}x \ \tilde{N}\left[\tilde{N}_{z}\partial_{\nu}\tilde{N}^{\nu}-\frac{1}{2}\tilde{N}_{\nu}(\partial_{z}\tilde{N}^{\nu}+\partial^{\nu}\tilde{N}_{z})\right],
\end{eqnarray}
where we have been careful to keep surface terms.  Plugging in (\ref{piconfig2}),
\begin{eqnarray}
S_{\mathcal{V}}&=&\int_{\mathcal{V}} d^nX \ \partial_{\mu}\partial_{y}\Pi\left[\partial^{\mu}\Pi\square\Pi -\partial_{\nu}\Pi\partial^{\mu}\partial^{\nu}\Pi\right]\nonumber\\
&&-\int_{\mathcal{B}_{z}}d^{n-1}x \ \partial_{y}\Pi\left[\partial_{z}\Pi\square\Pi -\partial_{\nu}\Pi\partial_{z}\partial^{\nu}\Pi\right]\nonumber\\
&=&\int_{\mathcal{V}} d^nX \ \partial_{y}\left[\frac{1}{2}(\partial_{\nu}\Pi)^{2}\square\Pi\right]-\partial_{\mu}\left[\frac{1}{2}(\partial_{\nu}\Pi)^{2}\partial_{y}\partial^{\mu}\Pi\right]\nonumber\\
&&-\int_{\mathcal{B}_{z}}d^{n-1}x \ \partial_{y}\Pi\left[\partial_{z}\Pi\square\Pi -\partial_{\nu}\Pi\partial_{z}\partial^{\nu}\Pi\right]\nonumber\\
&=&\int_{\mathcal{V}} d^nX \ \partial_{y}\left[\frac{1}{2}(\partial_{\nu}\Pi)^{2}\square\Pi\right]\nonumber\\
&&-\int_{\mathcal{B}_{z}}d^{n-1}x \ \partial_{y}\Pi(\partial_{z}\Pi\partial_{i}^2\Pi -\partial_{i}\Pi\partial_{z}\partial_{i}\Pi)+\frac{1}{2}(\partial_{\nu}\Pi)^{2}\partial_{y}\partial_{z}\Pi\nonumber\\
&=&\int_{\mathcal{V}} d^nX \ \partial_{y}\left[\frac{1}{2}(\partial_{\nu}\Pi)^{2}\square\Pi\right]\nonumber\\
&&-\int_{\mathcal{B}_{z}}d^{n-1}x \ \partial_{y}\left[\frac{1}{6}(\partial_{z}\Pi)^{3}+\frac{1}{2}(\partial_{i}\Pi)^{2}\partial_{z}\Pi\right] +2\partial_{y}\Pi\partial_{z}\Pi\partial_{i}^2\Pi.\nn\\ \label{unwanted}
\end{eqnarray}
Note, the volume integral gives precisely the $\pi$-Lagrangian once the $y$ integral is done.  The surface term gives the correct boundary term (\ref{piboundary2}) after integrating $y$, plus an unwanted piece $2\partial_{y}\Pi\partial_{z}\Pi\partial_{i}^2\Pi$, which is not even a total $y$ derivative.  It is this piece that will be cancelled off by the contributions from the Gibbons-Hawking-York and corner terms.  

\subsection*{Boundary piece, $S_{{\cal B}_z}$}
\ \ \ \ \
We now start with the terms localized on ${\cal B}_z$,
\begin{eqnarray}
S_{\mathcal{B}_{z}}=2\int_{\mathcal{B}_{z}}d^{n-1}x\sqrt{-\gamma}\left[g^{\mu\nu}e^{A}_{\mu}e^{B}_{\nu}\nabla_{B}z_{A} + n^{B}\nabla_{B}(n^{A}z_{A})-n^{A}z_{A}K\right].\nonumber\\
\end{eqnarray}
Using the expressions (\ref{expressions}), we calculate,
\begin{eqnarray}
\sqrt{-\gamma} g^{\mu\nu}e^{A}_{\mu}e^{B}_{\nu}\nabla_{B}z_{A}&=&\tilde{N}_{z}^{2}\partial_{z}\tilde{N}-\tilde{N}\tilde{N}_{z}\partial_{i}\tilde{N}_{i},\\
\sqrt{-\gamma}n^{B}\nabla_{B}(n^{A}z_{A})&=&\partial_{y}[\frac{1}{2}\tilde{N}_{z}^{3}-\tilde{N}_{z}\tilde{N}^{2}]-\tilde{N}^{\mu}\partial_{\mu}(\tilde{N}\tilde{N}_{z})\nn \\ && -\frac{1}{2}\tilde{N}_{z}^{2}\partial_{y}(\tilde{N}_{z}),\\
\sqrt{-\gamma}n^{A}z_{A}K&=&-\tilde{N}\tilde{N}_{z}\partial_{\mu}\tilde{N}^{\mu}.
\end{eqnarray}
Combining these expressions we have,
\begin{eqnarray}\label{expression8}
S_{\mathcal{B}_{z}}&=&2\int_{\mathcal{B}_{z}}d^{n-1}x\left[-\tilde{N}_{i}\partial_{i}(\tilde{N}\tilde{N}_{z}) +\partial_{y}[\frac{1}{2}\tilde{N}_{z}^{3}-\tilde{N}_{z}\tilde{N}^{2}-\frac{1}{6}\tilde{N}_{z}^{3}]\right].\nonumber\\
\end{eqnarray}
The first term becomes, 
\be -2\int_{\mathcal{B}_{z}}d^{n-1}x\ \partial_i\Pi\partial_{i}\left(\partial_y\Pi\partial_{z}\Pi\right),\ee
which after integrating $\partial_i$ by parts cancels the unwanted term in (\ref{unwanted}).  The remaining terms in (\ref{expression8}) will be canceled by the corner term.  

\subsection*{Corner piece, $S_{\rm corner}$}
\ \ \ \ \
Lastly, the corner term must be included,
\begin{eqnarray}
S_{\mathcal{B}_{y}\cap\mathcal{B}_{z}}=-2\int_{\mathcal{B}_{y}\cap\mathcal{B}_{z}}d^{n-2}x\sqrt{-\sigma} \arccos(n^{A}z_{A}). \nonumber
\end{eqnarray}
Expanding to third order,
\begin{eqnarray}
\sqrt{-\sigma} \arccos(n^{A}z_{A}) &=&- n^{A}z_{A} - \frac{1}{6}(n^{A}z_{A})^{3}\nonumber\\
&=&-\frac{1}{2}\tilde{N}_{z}^{3}+\tilde{N}_{z}\tilde{N}^{2}+\frac{1}{6}\tilde{N}_{z}^{3},
\end{eqnarray}
where we have dropped a constant term.  

Finally, adding all three terms together and performing the y integration, the full action at third order becomes,
\begin{eqnarray}
S=-\frac{1}{2}\int_{\mathcal{B}_{y}}d^{n-1}x(\partial_{\mu}\pi)^{2}\square\pi +\int_{\mathcal{B}_{y}\cap\mathcal{B}_{z}}d^{n-2}x\ \frac{1}{6}(\partial_{z}\pi)^{3} +\frac{1}{2}(\partial_{i}\pi)^{2}\partial_{z}\pi,
\end{eqnarray}
reproducing the familiar $\pi$-Lagrangian with the appropriate boundary terms (\ref{piboundary2}).

It should be clear at this point why the boundary ${\cal B}_y$ cannot be set orthogonal to ${\cal B}_z$ in general.  Making them orthogonal would be setting $N_z=0$, freezing to zero the degree of freedom in the normal derivatives of $\pi$,  the very part we are after.  

\section{\label{DGPjunction}An application to $\pi$-junction conditions}
\ \ \ \ \
In this section we make use of the new boundary terms and the variational principle to find junction conditions for $\pi$ across a sheet-like source, the analog of the Israel junction conditions of GR.  Consider the $\pi$ field in $d$-dimensions with coordinates $x^\mu=(x^i,z)$.  Suppose there is a $(d-1)$-dimensional sheet of matter at $z=0$, so that the stress tensor is localized to a delta function $T(x^i,z)=T(x^i)\delta(z)$.  The action includes terms on the left and right sides of the sheet, boundary terms on both sides of the sheet, and the coupling to $T(x^i)$ localized to the sheet (setting $\lambda=1$),
\bea \nn S&=&-\frac{1}{2}\left\{\int_{\rm left}d^dx+\int_{\rm right}d^dx\right\}\ (\partial_{\mu}\pi)^{2}\square\pi\\ \nn &&+\left\{\oint_{z=0^-}d^{d-1}x-\oint_{z=0^+}d^{d-1}x\right\}\ \left[\frac{1}{6}(\partial_{z}\pi)^{3}  +\frac{1}{2}(\partial_{i}\pi)^{2}\partial_{z}\pi\right]+\oint d^{d-1}x\ \pi T(x^i). \\
\eea
Varying the action, looking back at (\ref{pivary}), and keeping terms proportional to $\delta\pi$, we find
\bea\nn \delta S&=&\left\{\int_{\rm left}d^dx+\int_{\rm right}d^dx\right\}\left[(\square\pi)^{2}-(\partial_{\mu}\partial_\nu\pi)^{2}\right]\delta\pi \\ \nn
&& -2\left\{\oint_{z=0^-}d^{d-1}x-\oint_{z=0^+}d^{d-1}x\right\}\partial_i^2\pi\partial_z\pi\delta\pi+\oint d^{d-1}x\ T(x^i)\delta\pi. \\
\eea
Note that the variations of the boundary term not involving $\partial_z\delta \pi$ have contributed.  

In order for the action to be stationary, its variation must vanish for all $\delta\pi$, including those right on the sheet.  This implies (restoring $\lambda$),
\be \label{junctioncondition} 2\lambda\partial_i^2\pi\Delta\left[\partial_z\pi\right]=-T(x^i),\ee
where $\Delta\left[\ldots\right]\equiv \left[\ldots\right]_{z=0^+}-\left[\ldots\right]_{z=0^-}$ is the change in a quantity across the sheet.  We have assumed that $\pi$ is continuous across the sheet.  Thus the tangential derivatives $\partial_i\pi$ are also continuous, and the discontinuity lies only in the normal derivatives.  

The boundary terms make it simple to calculate the junction conditions, but the result can be checked directly from the equations of motion.  First note that the equations of motion can be written as a double total derivative,
\be (\square\pi)^{2}-(\partial_{\mu}\partial_\nu\pi)^{2}=\partial^\mu\partial^\nu\left[\partial_\mu\pi \partial_\nu \pi-\eta_{\mu\nu}(\partial\pi)^2\right]=-T(x^i)\delta(z).\ee
Now integrate both sides in $z$ across a small interval $-\epsilon<z<\epsilon$.  Do this at an arbitrary point $x^i$.  
\be \Delta\left[\partial^\nu\left(\partial_z\pi\partial_\nu\pi\right)-\partial_z\left(\partial\pi\right)^2\right]+\int_{-\epsilon}^\epsilon dz\ \partial^i\partial^\nu\left[\partial_i\pi\partial_\nu\pi-\eta_{i\nu}(\partial\pi)^2\right]=-T(x^i),\ee
\bea  \nn \Delta\left[\partial^i\left(\partial_z\pi\partial_i\pi\right)-\partial_z\left(\partial_i\pi\right)^2\right]+ \Delta\left[\partial^i\left(\partial_i\pi\partial_z\pi\right)\right]&&\\ +\int_{-\epsilon}^\epsilon dz\ \partial^i\partial^j\left[\partial_i\pi\partial_j\pi-\eta_{ij}(\partial\pi)^2\right]&=&-T(x^i).\eea
Within the remaining $z$ integral, there are first derivatives in $z$, but no higher derivatives in $z$.  $\pi$ is assumed continuous across the sheet, therefore its first $z$ derivative may be discontinuous but contains no delta function factors.  Only second and higher derivatives in $z$ may contain delta function factors.  Therefore the integral vanishes, and we are left with 
\be \Delta\left[\partial^i\left(\partial_z\pi\partial_i\pi\right)-\partial_z\left(\partial_i\pi\right)^2\right]+ \Delta\left[\partial^i\left(\partial_z\pi\partial_i\pi\right)\right]=2\partial_i^2\pi\Delta\left[\partial_z\pi\right]=-T(x^i),
\ee
reproducing (\ref{junctioncondition}).

\section{Boundary terms and junction conditions for the general galileon}
\ \ \ \ \
The DGP $\pi$-lagrangian has two important properties; its field equations are second order, and it is invariant up to a total derivative under the internal galilean transformations $\pi\rightarrow \pi+c+b_\mu x^\mu$, where $c,b_\mu$ are arbitrary real constants.  In \cite{Nicolis:2008in}, all possible lagrangians of a single scalar with these properties are classified in all dimensions.  These lagrangians are called galileons, and are interesting because they provide relatively well-behaved modifications of gravity.  Like the DGP $\pi$-lagrangian, the galileon lagrangians contain higher derivatives, yet their field equations are second order, so we expect boundary terms will be needed.  In this section, we calculate the boundary terms for all the galileon lagrangians, and the associated junction conditions.  

As shown in  \cite{Nicolis:2008in}, there is one galileon lagrangian at each order in $\pi$.  Here, order refers to the number of $\pi$'
s that appear in the equations of motion, i.e. the DGP $\pi$ lagrangian will be the second order galileon.   The $n$-th order galileon is
\bea \nn {\cal L}_n=\eta^{\mu_1\nu_1\mu_2\nu_2\cdots\mu_n\nu_n}&&\left( \partial_{\mu_1}\pi\partial_{\nu_1}\pi\partial_{\mu_2}\partial_{\nu_2}\pi\cdots\partial_{\mu_n}\partial_{\nu_n}\pi\right.  \\ \nn && +\partial_{\mu_1}\partial_{\nu_1}\pi\partial_{\mu_2}\pi\partial_{\nu_2}\pi\cdots\partial_{\mu_n}\partial_{\nu_n}\pi \\ \nn &&+\cdots \\ \nn &&\left.+\partial_{\mu_1}\partial_{\nu_1}\pi\partial_{\mu_2}\partial_{\nu_2}\pi\cdots\partial_{\mu_n}\pi\partial_{\nu_n}\pi\right), \\ \label{galileon}
\eea
where 
\be \label{tensor} \eta^{\mu_1\nu_1\mu_2\nu_2\cdots\mu_n\nu_n}\equiv{1\over n!}\sum_p\left(-1\right)^{p}\eta^{\mu_1p(\nu_1)}\eta^{\mu_2p(\nu_2)}\cdots\eta^{\mu_np(\nu_n)}.
\ee 
The sum in (\ref{tensor}) is over all permutations of the $\nu$ indices and $(-1)^p$ is the sign of the permutation.  The tensor (\ref{tensor}) is anti-symmetric in the $\mu$'s, anti-symmetric the $\nu$'s, and symmetric under interchange of any $\mu,\nu$ pair with another.  Using the symmetry of (\ref{tensor}) under interchange of $\mu,\nu$ pairs, the lagrangian (\ref{galileon}) can also be written
\be\label{galileon2} {\cal L}_n=n\eta^{\mu_1\nu_1\mu_2\nu_2\cdots\mu_n\nu_n}\left( \partial_{\mu_1}\pi\partial_{\nu_1}\pi\partial_{\mu_2}\partial_{\nu_2}\pi\cdots\partial_{\mu_n}\partial_{\nu_n}\pi\right).
\ee 
These lagrangians are unique up to total derivatives and overall constants.   In $n$-dimensions, only the first $n$ galileons are non-trivial, i.e. not total derivatives.  

At first order, we have 
\be {\cal L}_1=\left(\partial\pi\right)^2,\ee
which is the standard kinetic term for a scalar.  At second order, we have the DGP $\pi$-lagrangian
\be {\cal L}_2=\left(\partial\pi\right)^2\square\pi-\partial_{\mu}\partial_{\nu}\pi\partial^\mu\pi\partial^\nu\pi,\ee
here in a form which differs from (\ref{piLagrangian2}) by a total derivative (and an overall constant).  Of course changing the lagrangian by a total derivative will change the necessary boundary term, since a total derivative is a boundary contribution.  We will find the boundary term for the lagrangian (\ref{galileon}), and a lagrangian that differs by a total derivative (such as (\ref{piLagrangian2})) will need to have its boundary term modified accordingly.  

Unlike the $n=2$ case, for $n>2$ there is no known higher dimensional gravitational setup such as DGP that yields the galileon in a decoupling limit.  As such, we will only be able to derive the boundary terms from consistency of the variational principle, not from a higher dimensional well-posed action as we did from DGP in the $n=2$ case.  

The equations of motion derived from (\ref{galileon}) are
\be \label{galileoneom} {\cal E}_n={\delta{\cal L}\over \delta \pi}=-n(n+1)\eta^{\mu_1\nu_1\mu_2\nu_2\cdots\mu_n\nu_n}\left( \partial_{\mu_1}\partial_{\nu_1}\pi\partial_{\mu_2}\partial_{\nu_2}\pi\cdots\partial_{\mu_n}\partial_{\nu_n}\pi\right),\ee
and are second order, as advertised.  


\subsection*{Boundary terms}
\ \ \ \ \
Varying the action (\ref{galileon2}), we find the following boundary contribution, remembering that $\delta\pi$ and all its tangential derivatives vanish on the boundary, and using the symmetry of (\ref{tensor}) under interchange of $\mu,\nu$ pairs,
\be \label{expression} \left.\delta{\cal L}_n\right|_{{\cal B}_z}=n(n-1)\eta^{zz\mu_2\nu_2\cdots\mu_n\nu_n}\partial_z\delta\pi \left(\partial_{\mu_2}\pi\partial_{\nu_2}\pi\partial_{\mu_3}\partial_{\nu_3}\pi\cdots\partial_{\mu_n}\partial_{\nu_n}\pi\right).\ee
Now, by the anti-symmetry of (\ref{tensor}) in the $\mu$'s and $\nu$'s, all the remaining $\mu,\nu$ indices in (\ref{expression}) must lie tangent to the boundary, so we have 
\be  \left.\delta{\cal L}_n\right|_{{\cal B}_z}=n(n-1)\eta^{zzi_2j_2\cdots i_nj_n}\partial_z\delta\pi \left(\partial_{i_2}\pi\partial_{j_2}\pi\partial_{i_3}\partial_{j_3}\pi\cdots\partial_{i_n}\partial_{j_n}\pi\right),\ee
where $i,j$ indices are along the boundary.  Using the property 
\be \label{lowerdimproperty} \eta^{zzi_2j_2\cdots i_nj_n}={1\over n}\eta^{i_2j_2\cdots i_nj_n},\ee
we have 
\be \left.\delta{\cal L}_n\right|_{{\cal B}_z}=\partial_z\delta\pi(n-1)\eta^{i_1j_1\cdots i_{n-1}j_{n-1}} \left(\partial_{i_1}\pi\partial_{j_1}\pi\partial_{i_2}\partial_{j_2}\pi\cdots\partial_{i_{n-1}}\partial_{j_{n-1}}\pi\right)=\partial_z\delta\pi \left.{\cal L}_{n-1}\right|_{{\cal B}_z},\ee
where $\left.{\cal L}_{n-1}\right|_{{\cal B}_z}$ is the $(n-1)$-th order galileon action using only boundary derivatives.  

The boundary term needed to cancel this variation is, up to a total tangential derivative and terms involving no normal derivatives,
\be \label{galileonboundary}\left({\cal L}_n\right)_{{\cal B}_z}=-\partial_z\pi \left.{\cal L}_{n-1}\right|_{{\cal B}_z}.\ee
Interestingly, the boundary term for the $n$-th order galileon is simply the $(n-1)$-th order galileon, times a normal derivative of $\pi$.  Recall that in the case $n=2$ this boundary term differs from (\ref{piboundary2}) simply because the lagrangian ${\cal L}_2$ differed by a total derivative. 

\subsection*{Junction conditions}
\ \ \ \ \
We will derive the junction conditions for the general galileon both from the variational principle and directly from the equations of motion.  The setup and method are identical to the DGP case described in section \ref{DGPjunction}.  The galileon action in the presence of the junction is 
\bea \nn S&=&\left\{\int_{\rm left}d^dx+\int_{\rm right}d^dx\right\}\ {\cal L}_n \\ \nn &&+\left\{\oint_{z=0^-}d^{d-1}x-\oint_{z=0^+}d^{d-1}x\right\}\ \left({\cal L}_n\right)_{{\cal B}_z}+\oint d^{d-1}x\ \pi T(x^i). \\
\eea

We will show the variation of the action in detail, so for convenience in what follows, we define the following shorthand quantities:
\bea A&\equiv& \eta^{i_1j_1\ldots i_{n-1}j_{n-1}}\partial_z\pi \left(\partial_{i_1}\partial_{j_1}\pi\cdots\partial_{i_{n-1}}\partial_{j_{n-1}}\pi\right)\delta\pi, \\ 
B&\equiv&  \eta^{i_1j_1\ldots i_{n-1}j_{n-1}}\left(\partial_{i_1}\pi\ \partial_{j_1}\partial_z\pi\ \partial_{i_2}\partial_{j_2}\pi\cdots\partial_{i_{n-1}}\partial_{j_{n-1}}\pi\right)\delta\pi, \\ 
C&\equiv& \eta^{i_1j_1\ldots i_{n-1}j_{n-1}}\left(\partial_{j_1}\pi\ \partial_{i_1}\partial_z\pi\ \partial_{i_2}\partial_{j_2}\pi\cdots\partial_{i_{n-1}}\partial_{j_{n-1}}\pi\right)\delta\pi, \\ 
D&\equiv&  \eta^{i_1j_1\ldots i_{n-1}j_{n-1}}\left(\partial_{i_1}\pi\partial_{j_1}\pi\ \partial_{i_2}\partial_{j_2}\partial_z\pi\ \partial_{i_3}\partial_{j_3}\pi\cdots\partial_{i_{n-1}}\partial_{j_{n-1}}\pi\right)\delta\pi. \nn \\
\eea

We start by varying the boundary part $\left({\cal L}_n\right)_{{\cal B}_z}$, as given in (\ref{galileonboundary}),
\be\left({\cal L}_n\right)_{{\cal B}_z}=-(n-1) \eta^{i_1j_1\ldots i_{n-1}j_{n-1}}\partial_z\pi \left(\partial_{i_1}\pi\partial_{j_1}\pi\ \partial_{i_2}\partial_{j_2}\pi\cdots\partial_{i_{n-1}}\partial_{j_{n-1}}\pi\right).\ee 
The variation $\partial_z \delta\pi$ will be ignored because the boundary term was designed to have it cancel with the bulk.  The variation of the $\pi$'s with a singe derivative gives, upon integration by parts,
\be \label{singleds} 2(n-1)A+(n-1)(B+C).\ee
The variation of the double derivative $\pi$'s, upon integrating by parts once to remove the $i$ derivative from $\delta\pi$, gives,
\bea \nn (n-1)(n-2) \eta^{i_1j_1\ldots i_{n-1}j_{n-1}}&&\left(\partial_{i_2}\partial_z\pi\ \partial_{i_1}\pi\partial_{j_1}\pi\ \partial_{j_2}\delta \pi\ \partial_{i_3}\partial_{j_3}\pi\cdots\partial_{i_{n-1}}\partial_{j_{n-1}}\pi\ \right. \\ \nn
&&+\left. \partial_z\pi\ \partial_{i_1}\pi\ \partial_{i_2}\partial_{j_1}\pi\ \partial_{j_2}\delta\pi\ \partial_{i_3}\partial_{j_3}\pi\cdots\partial_{i_{n-1}}\partial_{j_{n-1}}\pi\ \right). \\
\eea
Now another integration by parts to remove the $j$ derivative from $\delta\pi$ gives, line for line,
\bea \nn &&-(n-1)(n-2)D+(n-1)(n-2)B \\ \nn
&& +(n-1)(n-2)A+(n-1)(n-2)C. \\ \label{doubleds}
\eea 
adding together (\ref{singleds}) and (\ref{doubleds}) gives the total variation of the boundary term
\be \label{boundaryvar}{ \delta \left({\cal L}_n\right)_{{\cal B}_z}=n(n-1)A+(n-1)^2(B+C)-(n-1)(n-2)D.}\ee

Now, start on the variation of the bulk term, as given in (\ref{galileon2}),
\be {\cal L}_n=n \eta^{\mu_1\nu_1\ldots\mu_n\nu_n}\left(\partial_{\mu_1}\pi\partial_{\nu_1}\pi\ \partial_{\mu_2}\partial_{\nu_2}\pi\cdots\partial_{\mu_{n}}\partial_{\nu_{n}}\pi\right).
\ee
We break the variation into three pieces:
\begin{itemize}
\item 
First consider the part of the variation on the boundary coming from varying the first derivative $\pi$'s and pulling off the derivative.  Call this $\mathbb{A}$, 
\bea  \nn \mathbb{A}=&& n \eta^{z\nu_1\ldots\mu_n\nu_n}\left(\delta\pi\ \partial_{\nu_1}\pi\ \partial_{\mu_2}\partial_{\nu_2}\pi\cdots\partial_{\mu_{n}}\partial_{\nu_{n}}\pi\right) \\
&+&n \eta^{\mu_1z\ldots\mu_n\nu_n}\left(\delta\pi\ \partial_{\mu_1}\pi\ \partial_{\mu_2}\partial_{\nu_2}\pi\cdots\partial_{\mu_{n}}\partial_{\nu_{n}}\pi\right).
\eea
Now, of the remaining $\mu,\nu$ indices, only one other at a time can take the value $z$, due to the antisymmetry of $\eta^{\mu_1\nu_1\ldots\mu_n\nu_n}$,
\bea  \nn  \mathbb{A}=&&n \eta^{zz\ i_2 j_2\ldots i_n j_n}\left(\delta\pi\ \partial_{z}\pi\ \partial_{i_2}\partial_{j_2}\pi\cdots\partial_{i_{n}}\partial_{j_{n}}\pi\right) \\ \nn
&+&n(n-1) \eta^{z j_1\ i_2 z\ \ldots i_n j_n}\left(\delta\pi\ \partial_{j_1}\pi\ \partial_{i_2}\partial_{z}\pi\ \partial_{i_3}\partial_{j_3}\pi\cdots\partial_{i_{n}}\partial_{j_{n}}\pi\right) \\ \nn \\
 \nn  &+&n \eta^{zz\ i_2 j_2\ldots i_n j_n}\left(\delta\pi\ \partial_{z}\pi\ \partial_{i_2}\partial_{j_2}\pi\cdots\partial_{i_{n}}\partial_{j_{n}}\pi\right) \\ \nn
&+&n(n-1) \eta^{ i_1 z\ z j_2 \ \ldots i_n j_n}\left(\delta\pi\ \partial_{i_1}\pi\ \partial_{z}\partial_{j_2}\pi\ \partial_{i_3}\partial_{j_3}\pi\cdots\partial_{i_{n}}\partial_{j_{n}}\pi\right). \\
\eea
Using the property (\ref{lowerdimproperty}) and renaming indices,
\bea \nn  \mathbb{A}=&&A-(n-1)C \\ &+&A-(n-1)B, \\
 =&&2A-(n-1)(B+C).
\eea

\item 

Next consider the part of the variation on the boundary coming from varying the double derivative $\pi$'s and pulling off the $\mu$ derivative.  Call this $\mathbb{B}$, 
\be  \mathbb{B}= n(n-1) \eta^{\mu_1\nu_1\ z\nu_2\ \mu_3\nu_3\ldots\mu_n\nu_n}\left(\partial_{\mu_1}\pi\partial_{\nu_1}\pi\  \partial_{\nu_2}\delta\pi\ \partial_{\mu_3}\partial_{\nu_3}\pi\cdots\partial_{\mu_{n}}\partial_{\nu_{n}}\pi\right). 
\ee
We can set $\nu_2=j_2$, because as mentioned before the variation $\partial_z \delta\pi$ cancels with the boundary term,
\be  \mathbb{B}= n(n-1) \eta^{\mu_1\nu_1\ z j_2\ \mu_3\nu_3\ldots\mu_n\nu_n}\left(\partial_{\mu_1}\pi\partial_{\nu_1}\pi\ \partial_{j_2}\delta\pi\ \partial_{\mu_3}\partial_{\nu_3}\pi\cdots\partial_{\mu_{n}}\partial_{\nu_{n}}\pi\right). 
\ee
Of the remaining $\mu,\nu$ indices, only one other at a time can take the value $z$, 
\bea  \nn  \mathbb{B}&=&n(n-1) \eta^{i_1 z\ z j_2\ i_3 j_3\ldots i_n j_n}\left(\partial_{i_1}\pi\ \partial_z\pi\ \partial_{j_2}\delta\pi\ \partial_{i_3}\partial_{j_3}\pi\cdots\partial_{i_{n}}\partial_{j_{n}}\pi\right) \\ \nn
&+&n(n-1)(n-2) \eta^{i_1 j_1\ z j_2\ i_3 z\ i_4 j_4\ldots i_n j_n}\left(\partial_{i_1}\pi\partial_{j_1}\pi\  \partial_{j_2}\delta\pi\ \partial_{i_3}\partial_z\pi\ \partial_{i_4}\partial_{j_4}\pi\cdots\partial_{i_{n}}\partial_{j_{n}}\pi\right). \\ 
\eea
Using the property (\ref{lowerdimproperty}) and renaming indices,
\bea \nn  \mathbb{B}=&-&(n-1)\eta^{i_1j_1\ldots i_{n-1}j_{n-1}}\left(\partial_{i_1}\pi\ \partial_z\pi\ \partial_{j_1}\delta\pi\ \partial_{i_2}\partial_{j_2}\pi\cdots\partial_{i_{n-1}}\partial_{j_{n-1}}\pi\right) \\ 
\nn &-& (n-1)(n-2) \eta^{i_1j_1\ldots i_{n-1}j_{n-1}}\left(\partial_{i_1}\pi\partial_{j_1}\pi\  \partial_{j_2}\delta\pi\ \partial_{i_2}\partial_z\pi\ \partial_{i_3}\partial_{j_3}\pi\cdots\partial_{i_{n}}\partial_{j_{n}}\pi\right), \\ 
\eea
and integrating by parts to pull the $j$ derivative off of $\delta\pi$, 
\bea \nn {\mathbb B}=&&(n-1)A+(n-1)B \\ &-&(n-1)(n-2)C+(n-1)(n-2)D.
\eea

\item

Finally, after pulling off the $\mu$ derivative from the double derivatives in the bulk term, there remains the bulk term
\be -n(n-1)\eta^{\mu_1\nu_1\ldots\mu_n\nu_n}\left(\partial_{\mu_1}\pi\ \partial_{\mu_2}\partial_{\nu_1}\pi\ \partial_{\nu_2}\delta\pi\ \partial_{\mu_3}\partial_{\nu_3}\pi\cdots\partial_{\mu_{n}}\partial_{\nu_{n}}\pi\right),\ee
which gives a boundary part which we label ${\mathbb C}$,
\be {\mathbb C}=-n(n-1)\eta^{\mu_1\nu_1\ \mu_2 z\ \mu_3\nu_3\ldots\mu_n\nu_n}\left(\partial_{\mu_1}\pi\ \partial_{\mu_2}\partial_{\nu_1}\pi\ \delta\pi\ \partial_{\mu_3}\partial_{\nu_3}\pi\cdots\partial_{\mu_{n}}\partial_{\nu_{n}}\pi\right)\ee

\bea \nn =&-& n(n-1) \eta^{z j_1 \ i_2 z \ i_3 j_3\ldots i_n j_n}\left(\partial_z\pi\ \partial_{i_2}\partial_{j_1}\pi\ \delta\pi\ \partial_{i_3}\partial_{j_3}\pi\cdots\partial_{i_{n}}\partial_{j_{n}}\pi\right) \\ \nn
&-&n(n-1)(n-2) \eta^{i_1 j_1\ i_2 z\ z  j_3\ i_4 j_4\ldots i_n j_n}\left(\partial_{i_1}\pi \ \partial_{i_2}\partial_{j_1}\pi\ \delta\pi\ \partial_z\partial_{j_3}\pi\ \partial_{i_4}\partial_{j_4}\pi\cdots\partial_{i_{n}}\partial_{j_{n}}\pi\right) \\ \nn
&-& n(n-1) \eta^{i_1 j_1\ zz \ i_3 j_3 \ \ldots i_n j_n}\left(\partial_{i_1}\pi \partial_z\partial_{j_1}\pi\ \delta\pi\ \partial_{i_3}\partial_{j_3}\pi\cdots\partial_{i_{n}}\partial_{j_{n}}\pi\right). \\ 
\eea
\bea\nn  {\mathbb C}&=&(n-1)A-(n-1)(n-2)B-(n-1)B\\ &=&(n-1)A-(n-1)^2B. \eea
\end{itemize}


In total, the bulk variation gives
\bea \left.\delta{\cal L}_n\right|_{\cal B}&=&{\mathbb A}+{\mathbb B}+{\mathbb C} \nn \\
\label{bulkvar}&=&2n A-(n-1)^2 (B+C)+(n-1)(n-2)D.\eea


Adding together the bulk and boundary variations (\ref{boundaryvar}) and (\ref{bulkvar}) on both sides of the boundary and demanding the action be stationary at the boundary leads to the discontinuity 
\be -\Delta\left[n(n+1)A\right]+T(x^i)=0.\ee
Using the assumption that $\pi$ and its tangential derivatives are continuous across the junction, we have finally
\be \label{galileonjunction} {n+1\over n-1}\Delta\left(\partial_z \pi\right)\left.{\cal E}_{n-1}\right|_{{\cal B}_z}=-T(x^i),\ee
where $\left.{\cal E}_{n-1}\right|_{{\cal B}_z}$ means the $(n-1)$-th order galileon equation of motion as given in (\ref{galileoneom}), using only boundary derivatives.  Interestingly, the $n$-th order junction condition is proportional to the $(n-1)$-th order equation of motion.  In the case $n=2$ this agrees with (\ref{junctioncondition}) after appropriate normalizations.  

We now re-derive this same result directly from the equations of motion, ${\cal E}_n=-T(x)$,
\be -n(n+1)\eta^{\mu_1\nu_1\mu_2\nu_2\cdots\mu_n\nu_n}\left( \partial_{\mu_1}\partial_{\nu_1}\pi\partial_{\mu_2}\partial_{\nu_2}\pi\cdots\partial_{\mu_n}\partial_{\nu_n}\pi\right)=-T(x^i)\delta(z).\ee
Integrate both sides in $z$ across a small interval $-\epsilon<z<\epsilon$, at an arbitrary point $x^i$,  
\be -n(n+1)\int_{-\epsilon}^\epsilon dz\ \eta^{\mu_1\nu_1\mu_2\nu_2\cdots\mu_n\nu_n}\left( \partial_{\mu_1}\partial_{\nu_1}\pi\partial_{\mu_2}\partial_{\nu_2}\pi\cdots\partial_{\mu_n}\partial_{\nu_n}\pi\right)=-T(x^i).\ee
In the sums over $\mu$'s and $\mu$'s we are only interested in those parts that contain double $z$ derivatives.  This is because $\pi$ is assumed continuous, so only the double derivatives can bring in delta function factors that contribute under the $z$-integral.  In addition, there can only be one double $z$ derivative at a time in a term, because of the anti-symmetry properties of $\eta^{\mu_1\nu_1\cdots}$, and there are $n$ such terms, all identical, due to the symmetry properties $\eta^{\mu_1\nu_1\cdots}$,
\be -n^2(n+1)\int_{-\epsilon}^\epsilon dz\ \eta^{zz i_2j_2\cdots i_nj_n}\left( \partial^2_z\pi \partial_{i_2}\partial_{j_2}\pi\cdots\partial_{i_n}\partial_{j_n}\pi\right)=-T(x^i).\ee
Now, integrate one of the $z$ derivatives by parts.  The resulting integral vanishes, because it contains no double $z$ derivatives, and what remains is the endpoints, which gives the discontinuity,
\be -n^2(n+1)\Delta\left[ \eta^{zz i_2j_2\cdots i_nj_n}\left( \partial_z\pi \partial_{i_2}\partial_{j_2}\pi\cdots\partial_{i_n}\partial_{j_n}\pi\right)\right]=-T(x^i),\ee
\be -n(n+1)\Delta\left(\partial_z \pi\right) \eta^{i_1 j_1\cdots i_{n-1}j_{n-1}}\left( \partial_{i_1}\partial_{j_1}\pi\cdots\partial_{i_{n-1}}\partial_{j_{n-1}}\pi\right)=-T(x^i),\ee
\be {n+1\over n-1}\Delta\left(\partial_z \pi\right)\left.{\cal E}_{n-1}\right|_{{\cal B}_z}=-T(x^i),\ee
reproducing (\ref{galileonjunction}).

\section{Conclusions and speculations}
\ \ \ \ \
We have shown that the DGP $\pi$-Lagrangian must be supplemented by a Gibbons-Hawking-York type boundary term and have calculated this term directly from the bulk.  In GR, the Gibbons-Hawking-York term plays a prominent role in the calculation of black hole entropy in the semi-classical approximation, among other things.  The partition function is calculated from the Euclidean action evaluated on a black hole configuration, and it is found that the entire contribution comes solely from the Gibbons-Hawking-York term on the boundary at infinity \cite{Brown:1992bq}.  

It might be that the boundary term we have found could be used to compute corrections to the entropy of a DGP black hole in the decoupling limit.  No exact DGP black hole solution is known, but if only the asymptotic behavior of $\pi$ in the appropriate Euclidean solution is known, the boundary term could perhaps be evaluated.  However, because of no-hair theorems stating that a scalar such as $\pi$ must be trivial outside a black hole, it may very well be the case that there are no interesting corrections coming from $\pi$.  We leave these issues for the future.  

\bigskip
\goodbreak
\centerline{\bf Acknowledgements}
\noindent
\\
The authors are grateful to Lam Hui, Justin Khoury, Alberto Nicolis and Mark Trodden for discussions.  ED wishes to acknowledge Brian Greene for support through  DOE grant DE-FG02-92ER40699 and the Ohrstrom Foundation.  


\providecommand{\href}[2]{#2}\begingroup\raggedright\endgroup

\end{document}